\documentclass[final]{aipproc}

\layoutstyle{6x9}

\newcommand{\etal}{et\,al.\ }
\newcommand{\logg}{\mbox{$\log g$}}
\newcommand{\Teff}{\mbox{$T_\mathrm{eff}$}}

\newcommand{\ion}[2]{\mbox{#1\,{\sc #2}}}
\def\kpd{KPD\,0005$+$5106}

\def\elf{PG1159$-$035}
\def\pgvier{PG1424$+$535}

\begin{document}

\title{Discovery of Iron in PG1159 Stars}


\author{Klaus Werner}{
  address={Institute for Astronomy and Astrophysics, Kepler Center for
  Astro and Particle Physics, University
  of T\"ubingen, 72076 T\"ubingen, Germany}
}

\author{Thomas Rauch}{
  address={Institute for Astronomy and Astrophysics, Kepler Center for
  Astro and Particle Physics, University
  of T\"ubingen, 72076 T\"ubingen, Germany}
}

\author{Jeffrey~W.~Kruk}{
  address={NASA Goddard Space Flight Center, Greenbelt, MD 20771, USA}
}

\begin{abstract}
The lack of \ion{Fe}{vii} lines in PG1159 stars had led to the conclusion that
in some objects iron must be strongly depleted.  We have now detected \ion{Fe}{x}
lines in FUSE spectra of the very hottest PG1159 stars
($T\mathrm{\hspace*{-0.4ex}_{eff}}$\,=\,150\,000 -- 200\,000\,K;
RX\,J2117.1+3412, K\,1$-$16, NGC\,246, Longmore\,4). Surprisingly, we derive a
solar iron abundance. It is conspicuous that they are among the most massive
PG1159 stars (0.71 -- 0.82\,M\hspace*{0.2ex}$_\odot$), in contrast to those
objects for which strongest Fe-deficiency was claimed (0.53 --
0.56\,M\hspace*{0.2ex}$_\odot$).
Based on new \ion{Fe}{viii} line identifications in SOHO/SUMER UV spectra of the
Sun, we were able to detect these lines in FUSE spectra of several ``cooler''
(\Teff\,\raisebox{-0.4ex}{$\stackrel{<}{\scriptstyle
\sim} $}\,150\,000) objects, among them is the prototype \elf. An
abundance analysis is in progress.
\end{abstract}

\maketitle

\begin{figure}[th!]
\includegraphics[width=0.47\textwidth]{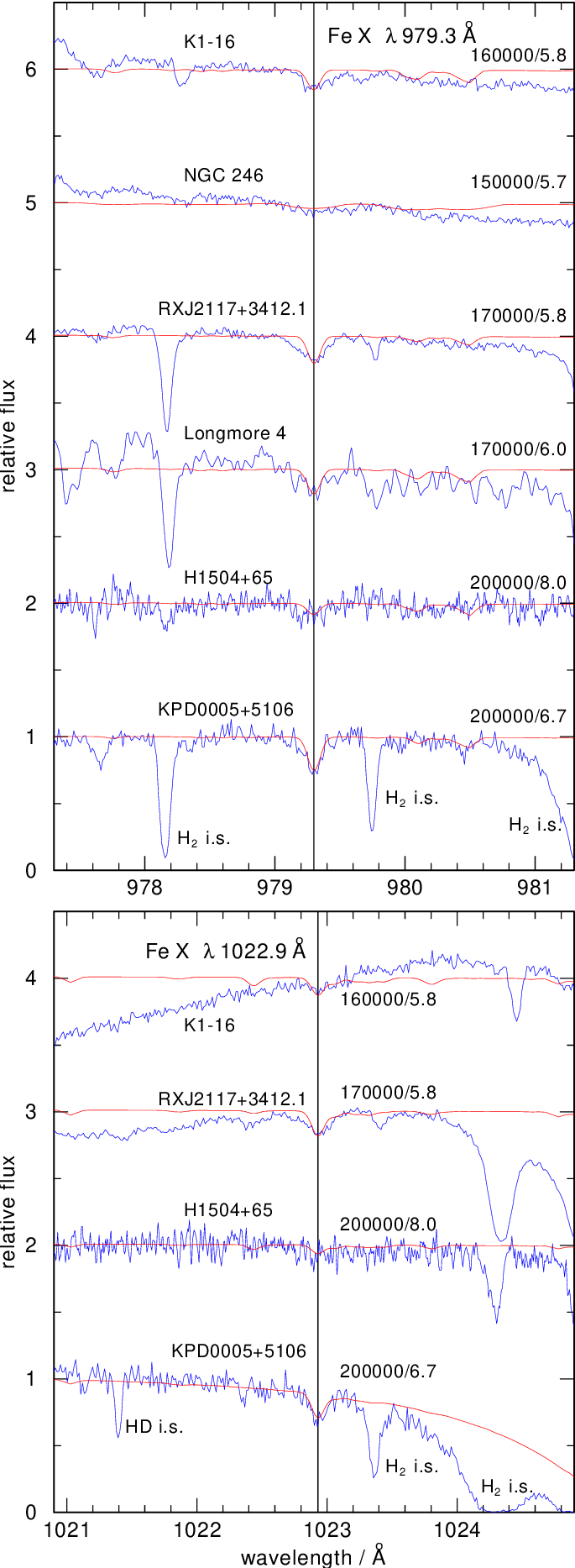}
\hspace{5mm}
\includegraphics[width=0.47\textwidth]{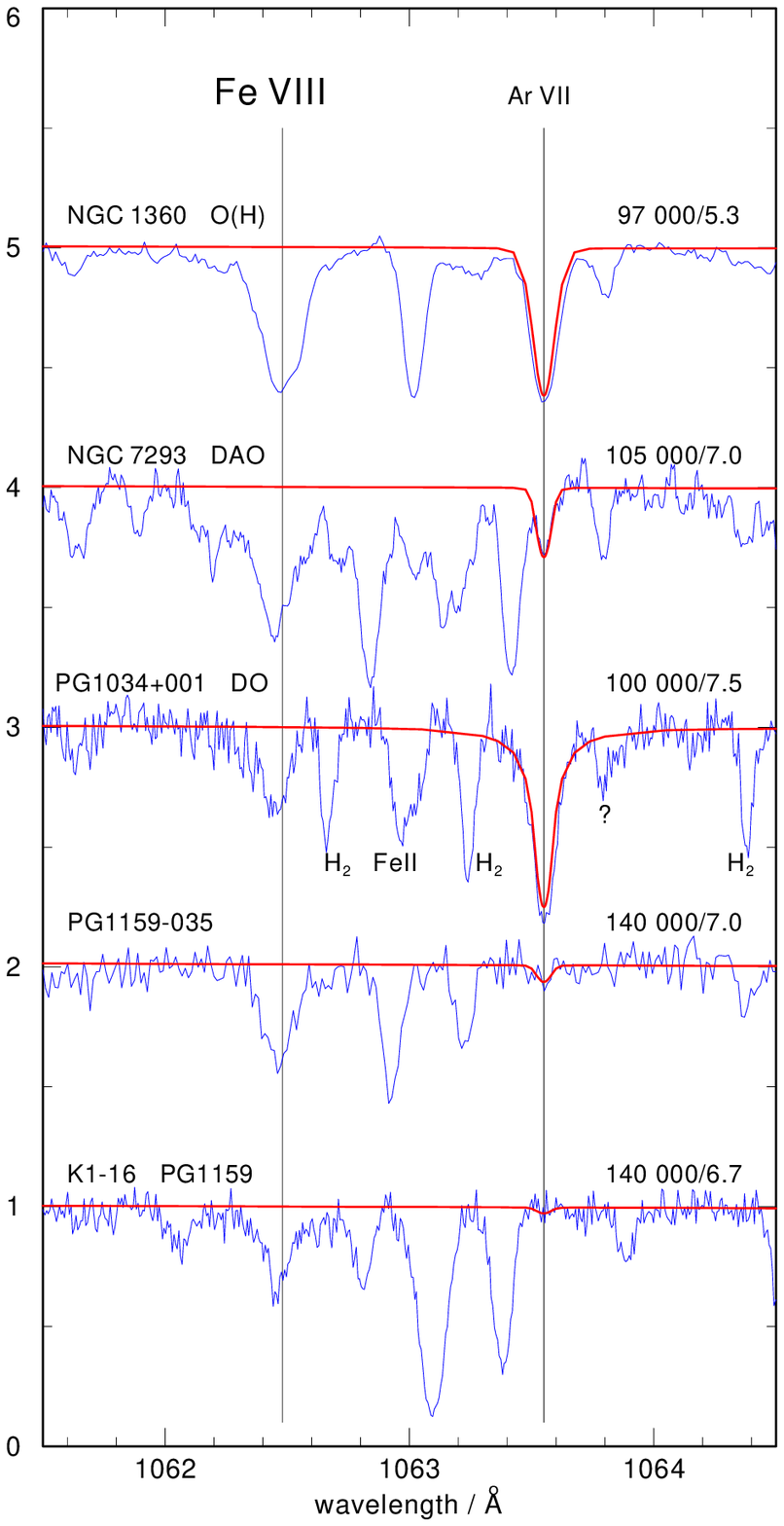}
\caption{\label{fig:fe}
\emph{Left:} Top panel: \ion{Fe}{x} $\lambda$979.3 in PG1159 stars and the
hot DO \kpd. Bottom: \ion{Fe}{x} $\lambda$1022.9 in PG1159 stars and
\kpd. Overplotted are models with solar Fe abundance. \Teff\ and \logg\
are indicated. \emph{Right:} \ion{Fe}{viii} $\lambda$1062.5,
located in the vicinity of an \ion{Ar}{vii} line, in a O(H)-type central star, a DO, a DAO, and two
PG1159 stars. The overplotted models include Ar but no Fe, yet.  }
\end{figure}                                                                                                                        

\section{Introduction}

Searching for iron in PG1159 stars poses a difficult problem. The atmospheric
parameter range (\Teff\,=\,75\,000 -- 200\,000\,K, \logg\,=\,5.5 -- 8.0
\cite{we06}) requires to look for lines from high ionization stages, at least
\ion{Fe}{vii}. Numerous \ion{Fe}{vii} lines are seen in UV spectra of hot
hydrogen-rich central stars of planetary nebulae (e.g., \cite{scho85}),  but the
search for these lines in PG1159 stars was unsuccessful. Model calculations show
that \ion{Fe}{vii} lines with detectable strength can only be expected in the
cooler PG1159 stars. They vanish with increasing \Teff\ because the ionization
balance of iron shifts to higher stages. The same tendency is seen with
decreasing gravity (i.e., increasing luminosity).

Based on the lack of \ion{Fe}{vii} lines, Fe-deficiency of $\geq 1$~dex was
concluded for four stars: \pgvier\ (\Teff\,=\,110\,000\,K, \logg\,=\,7.5
\cite{rei08}),  the hybrid-PG1159 stars (i.e., H-Balmer lines are visible)
NGC\,7094 and Abell\,43 (both have \Teff\,=\,100\,000\,K, \logg\,=\,5.5
\cite{zi09}), the PG1159 -- [WC] transition type star Abell\,78
(\Teff\,=\,110\,000\,K, \logg\,=\,5.5 \cite{we03}). Currently, these stars
represent the strongest cases for Fe-deficiency.  Also, Fe-deficiency was
claimed in some [WC]-type central stars, the putative progenitors of PG1159
stars. This finding is not explained by evolutionary models. We speculated that
Fe nuclei could have been transformed to heavier elements by excessive
s-processing.
                       
\begin{figure}[t]
\includegraphics[width=0.8\textwidth]{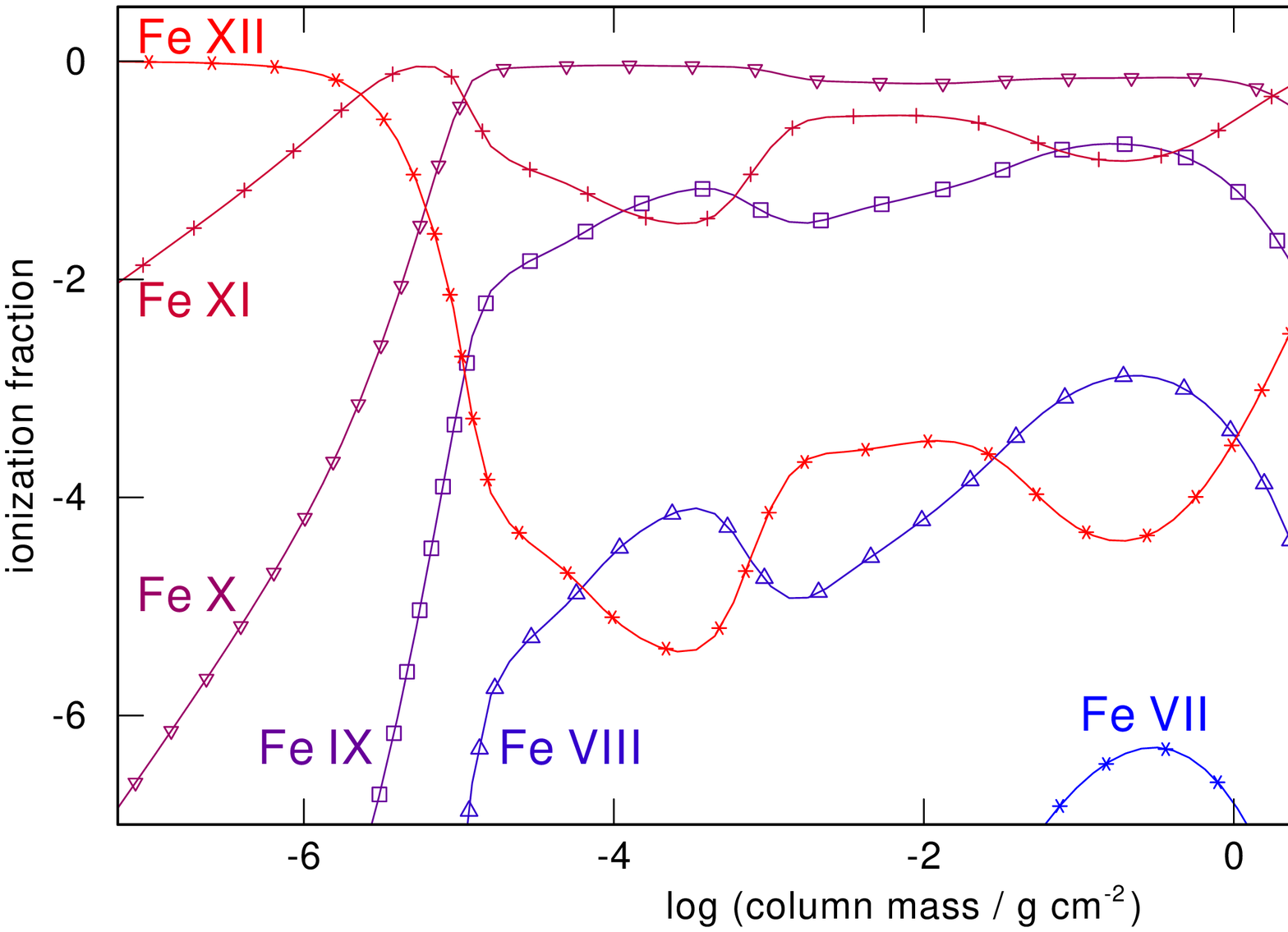}
\caption{\label{fig:ion}
Vertical run of iron ionization fractions in a model atmosphere with
\Teff=\,170\,000\,K and \logg=5.8.}
\end{figure}       

\begin{figure}[t]
\includegraphics[width=0.9\textwidth]{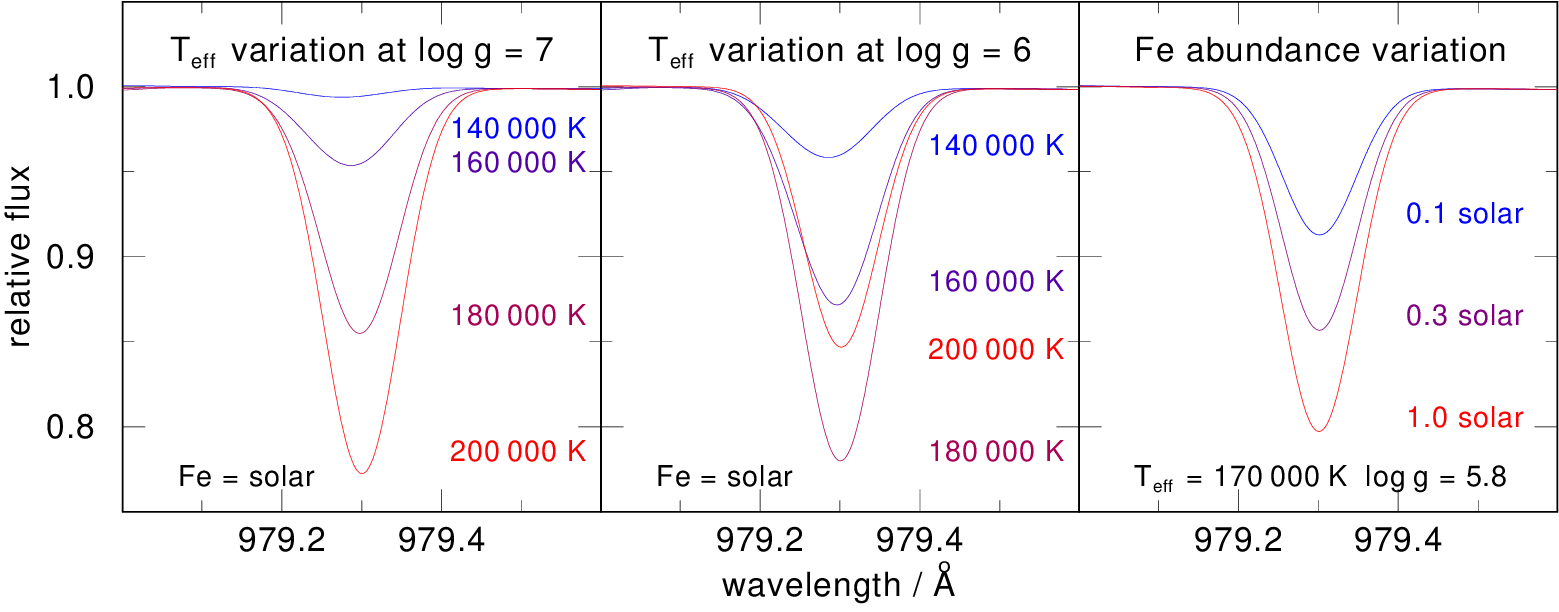}
\caption{\label{fig:var}
Dependence of \ion{Fe}{x} $\lambda$979.3 on
  atmospheric parameters. \emph{Left and middle panels:} \Teff\ variation at two
  fixed surface gravities.  \emph{Right panel:} variation
  of iron abundance at fixed \Teff\ and \logg.
}
\end{figure}       
                                                                                                                 
\section{Discovery of Fe\,X lines}

Two \ion{Fe}{x} lines were discovered in FUSE spectra of the very hottest PG1159
stars (Fig.\,\ref{fig:fe}, left panels). They are most prominent in the
high-luminosity central stars of K\,1--16, NGC\,246, RX\,J2117+3412.1, and
Longmore~4.  Fig.\,\ref{fig:ion} shows the ionization stratification of iron in
the model for RX\,J2117+3412.1. \ion{Fe}{x} is the dominant ionization stage
throughout the line-forming regions.

Iron abundances were determined by performing Fe NLTE calculations in our grid
of line-blanketed model atmospheres. Fig.\,\ref{fig:var} displays the behaviour
of one of the \ion{Fe}{x} lines when temperature, gravity, and abundance are
varied.  As a result of our analysis, we find solar iron abundances in the
hottest PG1159 stars \cite{we10}.  This contrasts with our above-mentioned
detections of Fe-deficiency in the four relatively cool PG1159 stars, based on
the absence of \ion{Fe}{vii} lines. At present, we do not see why the Fe
abundance determinations would be in error for the cooler stars; however we will
re-address this issue.

A scatter in Fe abundances among PG1159 stars would be difficult to
explain. However, there is one significant difference between the four putative
Fe-deficient stars and the four Fe-solar central stars. Comparison with
evolutionary tracks \cite{al09} shows that the former have masses in the range
0.53 -- 0.56\,M{$_\odot$}, and the latter 0.71 -- 0.82\,M$_\odot$. Is this
correlation between Fe abundance and stellar mass relevant?

\begin{figure}[t]
\includegraphics[width=\textwidth]{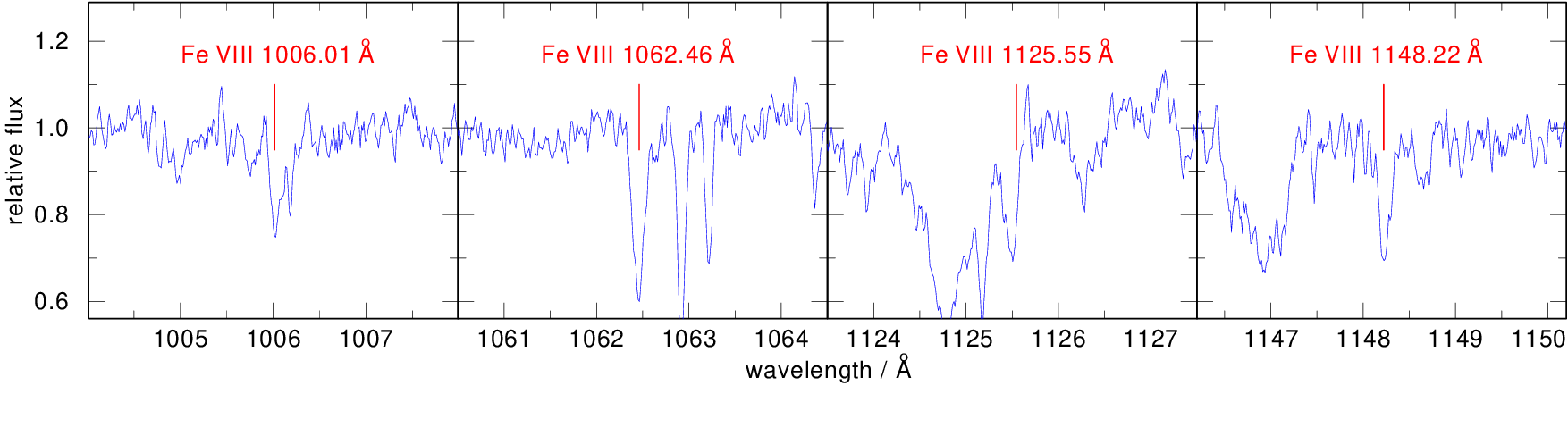}
\caption{\label{fig:pgelf}
Identification of four \ion{Fe}{viii} lines in \elf.
}
\end{figure}       

\section{Discovery of Fe\,VIII lines}

In a recent publication, Landi \& Young \cite{la10}  were able to identify four
\ion{Fe}{viii} emission lines in the $\lambda$1000 -- 1200 region of the quiet
Sun, in spectra obtained with the SOHO/SUMER instrument. This prompted us to
look for these lines in FUSE spectra of PG1159 stars. Indeed, they are seen in
several objects with \Teff\,=\,110\,000 -- 150\,000\,K. In particular, all four
\ion{Fe}{viii} lines are present in the prototype \elf\ (Fig.\,\ref{fig:pgelf}).
These detections will be an important tool to extend our iron abundance analyses
from the hottest objects to the more cooler stars.

Interestingly, these \ion{Fe}{viii} lines are also detected in spectra of
several of the hottest hydrogen-rich and DAO central stars, and hot DO white
dwarfs (Fig.\,\ref{fig:fe}, right panel). Together with previously identified \ion{Fe}{vii} lines,
they can be used as an independent temperature indicator.

\vspace{3mm}\noindent
{\bf Acknowledgements} \quad T.R. is supported by the German Aerospace Center
under grant 05\,OR\,0806.


\begin{thebibliography}{9}


\bibitem[{(){}}]{al09} Althaus, L. G.,  Panei, J. A., Miller
  Bertolami, M. M., \etal 2009, ApJ, 704, 1605

\bibitem[{(){}}]{la10} Landi, E., \& Young, P. R. 2010, ApJ, 713, 205

\bibitem[{(){}}]{rei08}  Reiff, E.,  Werner, K., Rauch, T., Koesterke, L., \& Kruk, J. W. 2008,
in Hydrogen-Deficient Stars, eds. K. Werner, T. Rauch, ASP Conf.
Ser., 391, 121

\bibitem[{(){}}]{scho85} Sch\"onberner, D., \& Drilling, J. S. 1985, ApJ, 290, L49

\bibitem[{(){}}]{we06} Werner, K., \& Herwig, F. 2006, PASP, 118, 183

\bibitem[{(){}}]{we03} Werner, K., Dreizler, S., Koesterke, L., \& Kruk, J. W. 2003,  
        in IAU Symp. 209: Planetary Nebulae,
        eds. S. Kwok, et al., Astronomical Society of the
        Pacific, p. 239

\bibitem[{(){}}]{we07} Werner, K., Rauch, T., \& Kruk, J. W. 2007, A\&A, 466, 317

\bibitem[{(){}}]{we10} Werner, K., Rauch, T., \& Kruk, J. W. 2010, ApJ, 719, L32

\bibitem[{(){}}]{zi09} Ziegler, M., Rauch, T., Werner, K., Koesterke, L., \& Kruk, J. W. 2009, 
in 16th European White Dwarf Workshop, eds. E. Garcia-Berro, et al., Journal of
Physics: Conf. Ser., 172, 012032


\end{thebibliography}
\end{document}